\documentstyle[11pt,aaspp4,graphics]{article}

\begin{document}

\newcommand{\lya}{Lyman-$\alpha$}
\newcommand{\eqw}{\hbox{EW}}
\def\erg{\hbox{erg}}
\def\cm{\hbox{cm}}
\def\sec{\hbox{s}}
\def\ha0{$6559$\AA}
\def\h16{$6730$\AA}
\def\f17{f_{17}}
\def\Mpc{\hbox{Mpc}}
\def\km{\hbox{km}}
\def\kms{\hbox{km s$^{-1}$}}
\def\year{\hbox{yr}}
\def\deg{\hbox{deg}}

% Stern's definitions:
% \def\lya{Ly$\alpha$}
\def\ergcm2s{\ifmmode {\rm\,erg\,cm^{-2}\,s^{-1}}\else
                ${\rm\,ergs\,cm^{-2}\,s^{-1}}$\fi}
\def\kmsMpc{\ifmmode {\rm\,km\,s^{-1}\,Mpc^{-1}}\else
                ${\rm\,km\,s^{-1}\,Mpc^{-1}}$\fi}
\def\nv{\ion{N}{5} $\lambda$1240}
\def\civ{\ion{C}{4} $\lambda$1549}
\def\oii{[\ion{O}{2}] $\lambda$3727}
\def\oiipair{[\ion{O}{2}] $\lambda \lambda$3726,3729}
\def\oiii{[\ion{O}{3}] $\lambda$5007}
\def\oiiipair{[\ion{O}{3}] $\lambda \lambda$4959,5007}

\title{First Results from the Large Area Lyman Alpha Survey}

\author{James E. Rhoads\altaffilmark{1}, Sangeeta Malhotra\altaffilmark{2,3},
 \& Arjun Dey\altaffilmark{2}}
\affil{Kitt Peak National Observatory}

\author{Daniel Stern\altaffilmark{4} \& Hyron Spinrad}
\affil{Astronomy Department, University of California,
    Berkeley, CA 94720}

\author{Buell T. Jannuzi}
\affil{National Optical Astronomy Observatories} 

\altaffiltext{1}{present address: Space Telescope Science Institute, 3700
San Martin Drive, Baltimore, MD 21218 ; rhoads@stsci.edu}
\altaffiltext{2}{Hubble Fellow}
\altaffiltext{3}{present address: Johns Hopkins University}
\altaffiltext{4}{present address: Jet Propulsion Laboratory, California
Institute of Technology, Mail Stop 169-327, Pasadena, CA 91109}  
\begin{abstract}
We report on a new survey for $z\approx 4.5$ \lya\ sources, the Large
Area Lyman Alpha (LALA) survey.  Our survey achieves an unprecedented
combination of volume and sensitivity by using narrow-band filters on
the new $8192^2$ pixel CCD Mosaic Camera at the 4 meter Mayall
telescope of Kitt Peak National Observatory.

Well-detected sources with flux and  equivalent width matching known
high redshift \lya\ galaxies (i.e., observed equivalent width $\eqw > 80$\AA,
$2.6 < \hbox{(line + continuum flux)}/ (10^{-17}\ergcm2s) < 5.2$,
and $\delta(\hbox{EW}) /
\hbox{EW} < 0.25$) have an observed surface density corresponding to $11000
\pm 700$ per square degree per unit redshift at $z=4.5$.  Spatial
variation in this surface density is apparent on comparison 
between counts in $6561 \pm 40$\AA\ and $6730 \pm 40$\AA\ filters.

Early spectroscopic followup results from the Keck telescope included
three sources meeting our criteria for good \lya\ candidates.  Of
these, one is confirmed as a $z=4.52$ source, while another remains
consistent with either $z=4.55$ or $z=0.81$.  We infer that $30$ to
$50\%$ of our good candidates are {\it bona fide\/} \lya\ emitters,
implying a net density of $\sim 4000 $ \lya\ emitters per square
degree per unit redshift.

% (redshift of confirmed source is 4.516)
\end{abstract}


\section{Introduction}

More than three decades ago Partridge and Peebles (1967) predicted
that galaxies in their first throes of star-formation should be strong
emitters in the \lya\ line. Their predictions were optimistic,
based on converting roughly 2\% of gas into stars in $3 \times 10^7$
years in Milky Way sized galaxies, which translates into a luminosity
of $6\times 10^{44} \erg\, \sec^{-1}$. These objects are also expected to be common -
if all the $L^*$ galaxies have undergone a phase of rapid star-formation
one should see a surface density of about $3 \times 10^3 \times
(\Delta t/(3 \times 10^7 \year)) \deg^{-2}$ (Pritchet 1994). Searches based
on these expectations did not detect \lya\  emitters (LAEs).
(See review by Pritchet 1994;  Koo \& Kron 1980; Pritchet \&
Hartwick 1987, 1990; Cowie 1988; Rhea et al 1989; Smith et al 1989;
Lowenthal et al 1990; Wolfe et al 1992; De Propris et al 1993;
Macchetto et al 1993; M{\o}ller \& Warren 1993; Djorgovski \& Thompson
1992; Djorgovski, Thompson, \& Smith 1993; Thompson, Djorgovski, \&
Trauger 1992; Thompson et al 1993; Thompson, Djorgovski, \& Beckwith
1994; Thommes et al 1998.)

Only recently have \lya\ emitters been observed, albeit at luminosity
levels roughly a hundred times lower than the original prediction.
These \lya\ emitters have been found from both deep narrow band
imaging surveys (Cowie \& Hu 1998; Hu, Cowie \& McMahon 1998; Hu,
McMahon, \& Cowie 1999; Kudritzki et al 2000), and from deep spectroscopic
surveys (Dey et al 1998; Manning et al 2000; but see Stern et al
2000).  Weak \lya\ emitters have also been found through targeted
spectroscopy of Lyman break objects (e.g., Steidel et al 1996,
Lowenthal et al 1997).  The lower luminosity in the \lya\ line may be
because of attenuation by dust if chemical enrichment is prompt; or
because the star-forming phase is more protracted; or because the
star-formation happens in smaller units which later merge. The first
two scenarios will give a smaller equivalent width than early
predictions, while the last scenario results in low luminosities but
high equivalent width.

Dust effects are expected to be severe--- even a small amount of dust
can greatly attenuate this line, because it is resonantly
scattered. However, two factors can help the \lya\ photons escape. If
\lya\ photons are produced in diffuse regions of a clumpy interstellar medium,
they can simply
scatter off the dense clumps and escape (Neufeld 1991), and some
geometries can even lead to an increase in the equivalent width of the
line. Secondly, energetic winds are seen in low-$z$ \lya\ emitters
(Kunth et al 1998).  These can displace the neutral gas and
doppler-shift the peak wavelength of the resonant scattering, thereby
reducing the amount of scattering and the path length for interaction
with dust.

Detailed predictions for luminosities and surface densities of LAEs
using a Press-Schechter formalism and exploring a range of dust
obscuration and star-formation time scales have been explored by
Haiman and Spaans (1999), who are able to reproduce
the surface densities of LAEs reported by Hu et al (1998) with a wide
range of models.  In order to narrow down the range of possibilities
and characterize the high redshift \lya\ population, better statistics
over a wide range of flux and source density are needed.

\section{Narrow Band Imaging Survey} \label{obsdesc}

An efficient search for \lya\ emitters (and other emission line
galaxies) was started in 1998 using the CCD Mosaic Camera at the Kitt
Peak National Observatory's 4m Mayall telescope. The Mosaic camera has
eight $2048 \times 4096$ chips in a $4 \times 2$ array comprising a
$36'\times 36'$ field of view. The final area covered by the LALA
survey is $0.72$ square-degrees in two MOSAIC fields centered at
14:25:57 +35:32 (2000.0) and 02:05:20 -04:55 (2000.0).
Five overlapping narrow band filters of width FWHM$\approx 80$\AA
are used. The central wavelengths are $6559$, $6611$, $6650$,
$6692$, and $6730 $\AA, giving a total redshift coverage $4.37 < z <
4.57$.  This translates into surveyed comoving volume of $8.5 \times
10^5$ comoving $\Mpc^3$ per field for $H_0 = 70
\kmsMpc$, $\Omega = 0.2$, $\Lambda=0$. About 70\% of the imaging
at $z \approx 4.5$ is complete, and an extension of the survey to
$z>5$ is planned. In about 6 hours per filter per field we are able to
achieve line detections of about $2 \times 10^{-17} \ergcm2s$. The
survey sensitivity varies with seeing.  Broadband images of these fields
in a custom $B_w$ filter ($\lambda_0 = 4135$\AA, $\hbox{FWHM}=1278
$\AA) and the Johnson-Cousins $R$, $I$, and $K$ bands are being taken
as part of the NOAO Deep Widefield Survey (Jannuzi \& Dey 1999).

The images were reduced using the MSCRED package (Valdes \& Tody 1998;
Valdes 1998) in the IRAF environment (Tody 1986, 1993), together with
assorted custom IRAF scripts.  In addition to standard CCD data
reduction steps (overscan subtraction, bias frame subtraction, and
flatfielding), it was necessary to remove crosstalk between pairs of
CCDs sharing readout electronics and to remove a ghost image of the
telescope pupil.  Astrometry of USNO-A catalog stars was used to
interpolate all chips and exposures onto a single coordinate grid
prior to combining final images.  Cosmic ray rejection is of
particular importance in a narrowband search for emission line
objects.  We therefore identified cosmic rays in individual images
using a digital filtering method (Rhoads 2000) and additionally
applied a sigma clipping algorithm at the image stacking stage.
Weights for each exposure were determined using the method of Fischer
\& Kochanski (1994), which accounts for sky level, transparency, and
seeing to optimize the signal to noise level for compact sources in
the final image.

Catalogs were generated using the SExtractor package % (v.2.0.2)
(Bertin \& Arnouts 1996).  Fluxes were measured in $2.32''$ (9 pixel) diameter
apertures, and colors were obtained using matched $2.32''$ apertures
in registered images that had been convolved to yield matched point spread
functions.

\section{Spectroscopic Observations}
Spectroscopic followup of a cross-section of emission line candidates
was obtained with the LRIS instrument (Oke et al 1995) at the Keck 10m telescope on
1999 June 13 (UT).  Two dithered exposures of $1800$ seconds each were obtained
through a single multislit mask in good weather.

These data were reduced using a combination of standard IRAF ONEDSPEC
and TWODSPEC tasks  together with the homegrown
slitmask reduction IRAF task ``BOGUS'' (Stern, Bunker, \& Stanford,
private communication) for reducing LRIS data.

\section{The emission line source population}
\begin{figure}[ht]
\plotone{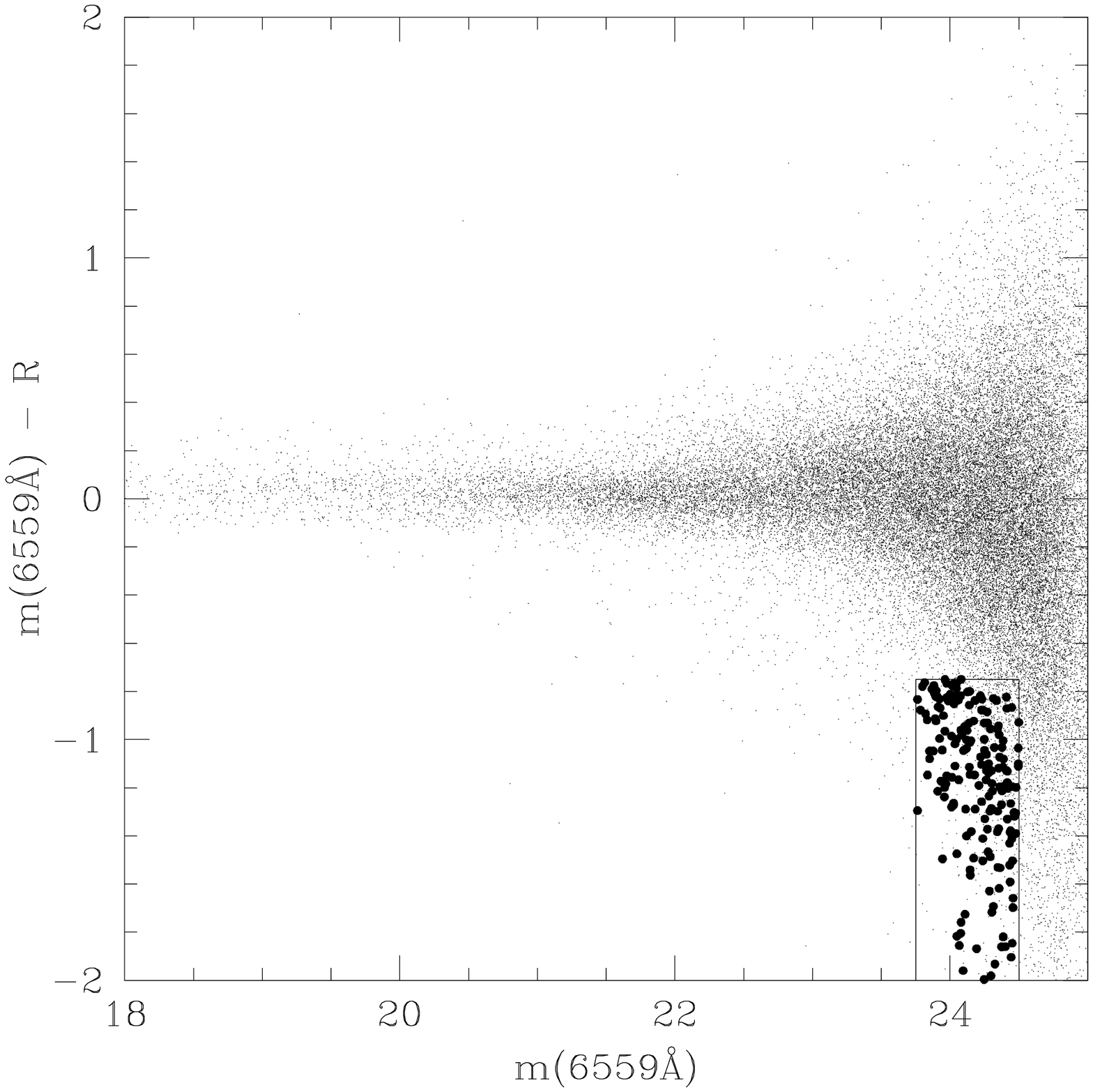}
\caption{
A sample LALA survey color-magnitude diagram, for narrowband \ha0\
magnitude (80\AA\ bandpass) and \ha0\ $-$ $R$ color.  The zero point of
the narrowband filter is calibrated using $R$ band standard star
fluxes (Landolt 1992), so that $m(6559\AA) = R$ for a pure continuum
source of average spectral slope.  The box in the lower right hand
corner shows the color and magnitude constraints for good \lya\
candidates, as described in the text.  Additional constraints on the
color error eliminate some objects meeting the color-magnitude cut;
the final sample of good candidates for this filter is marked by the
heavy points.
\label{cmdfig}}
\end{figure}

Our imaging data yield numbers of sources as a function of their
fluxes in several filters, from which we can construct number
densities as a function of magnitudes, colors, and equivalent widths.
In order to gracefully handle sources that are not detected in all
filters, we have chosen to use ``asinh magnitudes'' (Lupton, Gunn, \&
Szalay 1999), which are a logarithmic function of flux for well
detected sources but approach a linear function of flux for weak or
undetected sources.  Figure~\ref{cmdfig} shows the color-magnitude
diagram for the \ha0\ ($\pm 40 $\AA) and $R$ filters.  The color
scatter achieved for bright sources ($R<22$) is $0.10$ magnitudes
(semi-interquartile range).  This includes the true scatter in object
colors, and is therefore a firm upper limit on the scatter introduced
by any residual systematic error sources, which we expect to be a few
percent at worst.

To sharpen our focus on the high redshift \lya\ population, we
identify the range of parameter space occupied by known $z>3$ \lya\
emitters.  These sources have typical observed equivalent widths 
$\ga 80$\AA\ and line+continuum fluxes $\la 5 \times 10^{-17} \ergcm2s$
(adjusted to $z=4.5$ and measured in an $80$\AA\ filter).

We further want to restrict attention to sources with sufficiently
reliable detections that the number of false emission line candidates
is a small fraction of the total candidate sample.  The data set
includes many continuum sources ($\sim 1.4 \times 10^4$ in the flux
range above).  It is expected to contain a few hundred emission line
sources, based on earlier source counts from smaller samples (Hu et al
1998) and on our own findings (below).  The fraction of continuum
sources with measured equivalent widths above some threshold $\eqw_0$
can be calculated as a function of signal to noise level $n$ and
filter width $\Delta \lambda$: The measured equivalent width for a
source with no line emission or absorption in the narrowband filter
will be $0 \pm \Delta \lambda / n$.  Thus, a false positive becomes an
$m \sigma$ event, where $m = n \times \eqw_0 / \Delta \lambda$.  Our sample
would be reasonably safe from contamination with $m = 3$, which would
correspond to about 20 false positives in a sample of 14000 sources;
and very safe for $m \ga 4$, corresponding to $<1$ false positive in
14000 sources.  We have conservatively chosen detection thresholds
$n=5$ and $\eqw_0 = \Delta \lambda = 80$\AA, which gives $m=5$.  This
keeps the number of false positives small even when the foregoing
analysis is expanded to include errors in the continuum flux (which
increase photometric error in the color by a factor $\sim \sqrt{2}$)
and a realistic distribution of equivalent widths for the low-$z$
galaxy populations.  As a further check on our sample, we use the
photometric color error estimates from SExtractor to demand that the
source be an emission line source at the $4 \sigma$ level.

Combining all of these requirements, the final criteria for good
candidates in our survey become $\hbox{EW} > 80$\AA,
$\delta(\hbox{EW})/\hbox{EW} \le 0.25$, and $2.6 < \f17 < 5.2$, where
$\f17 \equiv f / (10^{-17} \ergcm2s)$.  There are $225$ such
sources detected in the \ha0\ filter in a solid angle of $0.31$ square
degree and redshift range $\delta z = 0.07$.  This corresponds to
$11000$ good candidate \lya\ emitters per square degree per unit
redshift.  The precise upper flux cutoff used is somewhat arbitrary,
but including sources with larger fluxes in our list of good
candidates has little effect on the total source counts.

The final \h16\ image had a broader PSF than the \ha0\ image, and
consequently a reduced effective sensitivity.  In addition, it samples
a slightly higher redshift, resulting in an $0.08$ magnitude increase
in distance modulus ($q_0=0.1$, $\Lambda=0$).  Accounting for both
effects, the matched flux range is $3.45 < \f17 < 5.2$ at \ha0\ and
$3.2 < \f17 < 4.8$ at \h16. The corresponding counts are $70$
candidates at \h16, and 104 candidates at \ha0.  Thus, the source
density in directly comparable luminosity bins varies by a factor of
$1.5$, a difference that is significant at about the $3\sigma$ level.
We interpret this as a likely signature of large scale structure in
the \lya\ emitter distribution at $z\approx 4.5$.  The comoving
distance between the centers of the two redshift slices is $72 \Mpc$,
while the thickness of each slice is $36 \Mpc$ and the transverse
comoving size $89 \Mpc$ (again assuming $H_0 = 70 \kmsMpc$,
$\Omega = 0.2$, $\Lambda=0$).  Comparable variations have been
observed in the density of Lyman break galaxies at $z \approx 3.1$
(Steidel et al 1998).

\section{Spectroscopic results}
\begin{figure}[ht]
\plotfiddle{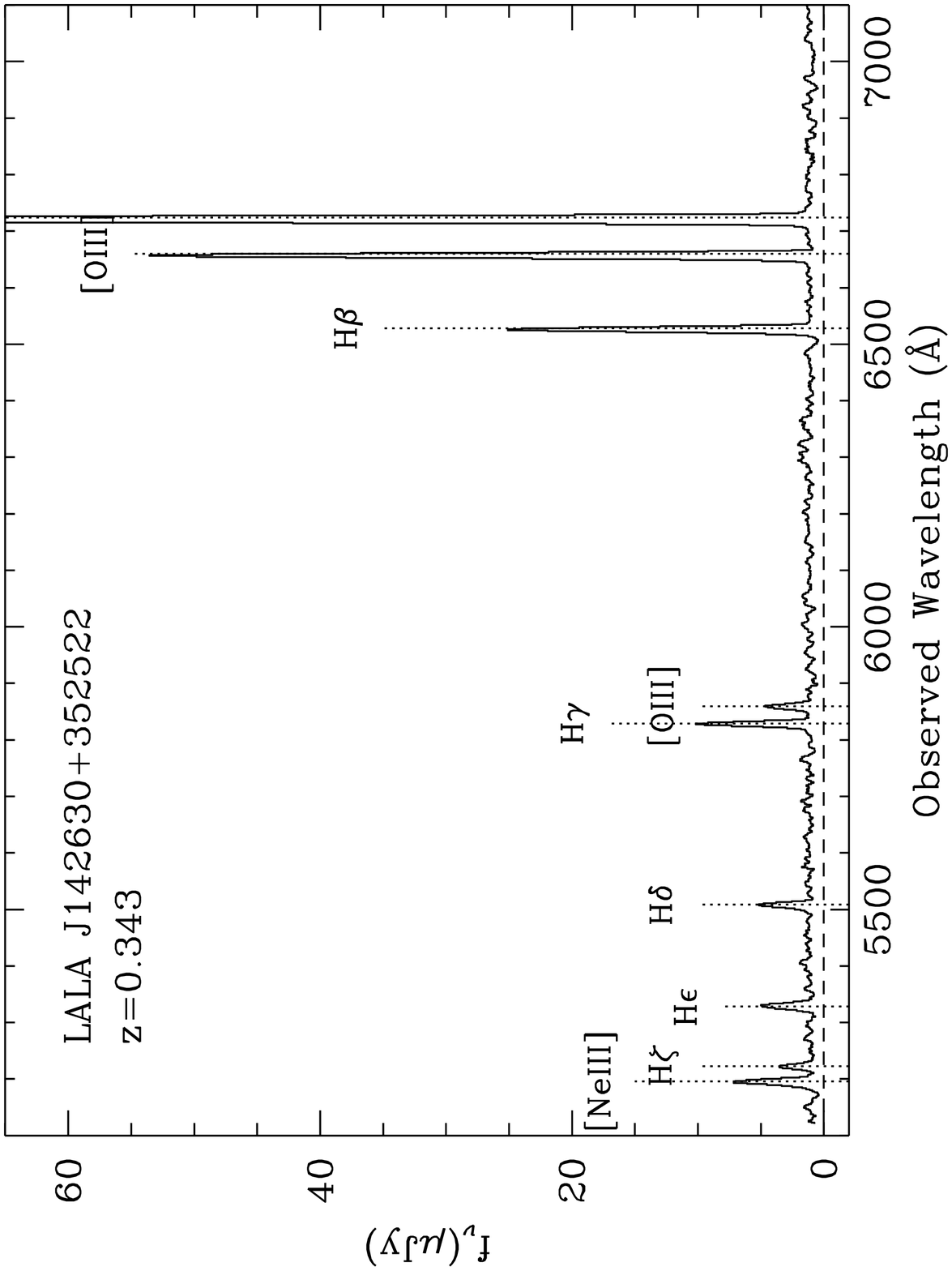}{3in}{-90}{42}{42}{-220}{265}
\plotfiddle{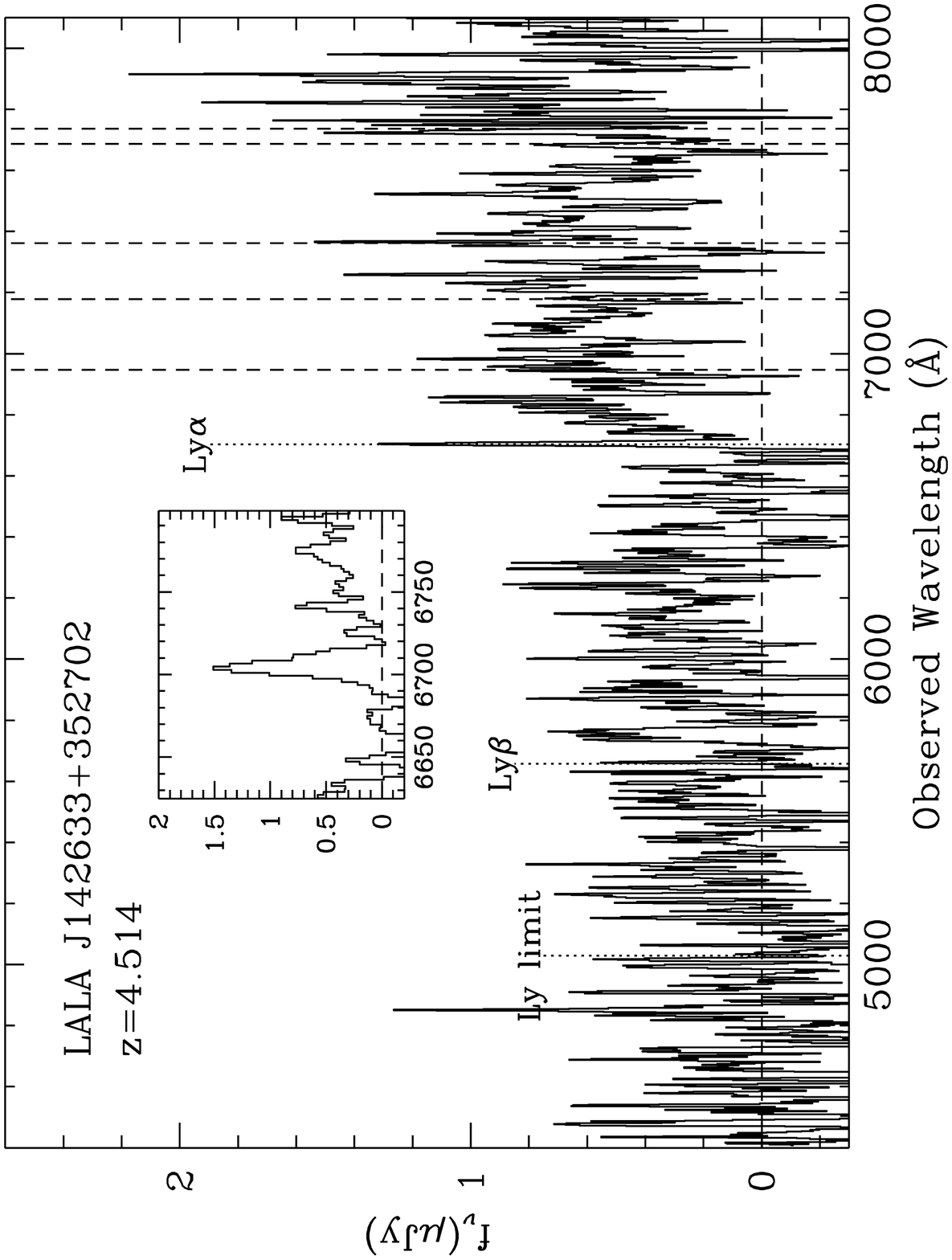}{3in}{-90}{42}{42}{-220}{255}
\caption{
Keck spectra of two LALA emission line candidates.  Top: The largest
equivalent width source in the spectroscopic sample, a $z=0.34$ OIII
emitter with observed $\eqw \approx 1400$\AA.  This source has eleven
detected emission lines, and an \oiii \AA\ line luminosity of
$2.5 \times 10^{41} \erg \sec^{-1}$.  It demonstrates our ability to find
unusual objects through our large volume coverage.  Bottom: A
confirmed $z=4.516$ \lya\ source.  This object has a line flux of $1.7
\times 10^{-17} \ergcm2s$ and an equivalent width of 84\AA.
The line is asymmetric and has a strong continuum decrement from the
red to the blue side, both of which are expected for high-redshift
\lya\ emitters (Stern \& Spinrad 1999).
It demonstrates our ability to find the faint \lya\ emitter population.
\label{specfig}}
\end{figure}

Our spectroscopic followup sampled a wide range of flux and equivalent
width in order to characterize the different populations of sources
with extreme narrowband colors, and to thereby tune candidate
selection criteria for future spectroscopic observations.  Those
sources with $5\sigma$ narrowband detections and photometrically
measured $\eqw > 65 $\AA\ were all confirmed as emission line objects
at the narrowband wavelength.  In total, we detected two \oiii\
emitters at $z=0.34$, three \oii\ emitters at
$z=0.77$ and $z=0.81$, one confirmed \lya\ ($1215$\AA) emitter at
$z=4.516$, and one source that could either be \lya\ at $z=4.55$ or
\oii at $z=0.81$.  We also found a $z=2.57$ galaxy with emission lines
of \civ, \ion{He}{2} $\lambda$1640, and \ion{O}{3} $\lambda$1663.  This
source may have been included in the narrowband candidate list (with
$\eqw=36$\AA) because of a weak spectral break around $6560$\AA.
Finally, we found a few serendipitous emission line sources
in the slit spectra.  One of these is a single-line source with large
equivalent width, possibly \lya\ at $z=3.99$.  
Spectra of two interesting sources are shown in figure~\ref{specfig}.

\section{Discussion: The \lya\ source population}
By combining our imaging survey with these spectroscopic results,
we can estimate the source density of \lya\ emitters passing our
selection cut.  Our spectra included three sources fulfilling the
criteria given above for good candidates.  Of these, one was confirmed
as a $z=4.52$ \lya\ source.  A second remains a candidate $z=4.55$
source, but is more conservatively interpreted as a $z=0.81$ [O II]
emitter on the basis of a rather strong continuum on the blue side of
the line.  The third is a clear $z=0.34$ [O III] emitter.
We therefore estimate that roughly  $1/3$ to $1/2$ of the
good candidates will be confirmed as \lya\ sources, yielding $\sim
4000$ emitters per square degree per unit redshift.  This is
compatible with earlier measurements from smaller volumes  (Hu et al
1998) after accounting for differences in flux threshold.

Our measurement is distinct from previous efforts in the field for its
basis in a large number of candidate emitters.  Poisson errors in our
source counts are of order $\pm 7\%$.  This is smaller than the
variations observed in the comparison of two filters (of order $\pm
40\%$).  By combining observations in multiple fields, we will be able
to average over local fluctuations in number densities effectively.
When completed, the LALA survey will yield comoving volume of $\sim 2
\times 10^6 \Mpc^3$ (\S \ref{obsdesc}) and a sample of several hundred
LAEs, and will allow the luminosity function, equivalent width
distribution, and correlation function of this population to be
determined for the first time.

\acknowledgements
We thank Andy Bunker and Steve Dawson for help with the spectroscopic
observations and Frank Valdes for writing and helping with the MSCRED
package in IRAF.  JER's research is supported by a Kitt Peak
Postdoctoral Fellowship and by an STScI Institute Fellowship.  SM's
research is supported by NASA through Hubble Fellowship grant \#
HF-01111.01-98A from the Space Telescope Science Institute, which is
operated by the Association of Universities for Research in Astronomy,
Inc., under NASA contract NAS5-26555.

\end{document}